\documentclass[twocolumn,epjc3]{svjour3}  
\smartqed  
\RequirePackage{graphicx}
\journalname{Eur. Phys. J. C}
\def\beq{\begin{equation}}
\def\eeq{\end{equation}}

\begin{document}

\title{Friction forces in cosmological models
}

\author{Donato Bini\thanksref{e1,addr1,addr2,infn,inaf}
\and
        Andrea Geralico\thanksref{e2,addr2,addr3} 
\and
        Daniele Gregoris\thanksref{e3,addr4,addr5,addr6}
\and
Sauro Succi\thanksref{e4,addr1,infn}
}

\thankstext{e1}{e-mail: binid@icra.it}
\thankstext{e2}{e-mail: geralico@icra.it}
\thankstext{e3}{e-mail: danielegregoris@libero.it}
\thankstext{e4}{e-mail: s.succi@iac.cnr.it}

\institute{Istituto per le Applicazioni del Calcolo \lq\lq M. Picone," CNR, I-00185 Rome, Italy\label{addr1}
           \and
           ICRA, University of Rome \lq\lq La Sapienza," I-00185 Rome, Italy\label{addr2}
           \and
           INFN, Sezione di Firenze, I--00185, Sesto Fiorentino (FI), Italy\label{infn}
           \and
           INAF, Astronomical Observatory of Torino, I--10025 Pino Torinese (TO), Italy\label{inaf}
           \and
           Physics Department, University of Rome \lq\lq La Sapienza," I-00185 Rome, Italy\label{addr3}
           \and
           Physics Department, University of Stockholm, SE-106 91 Stockholm, Sweden\label{addr4}
\and
Max-Planck-Institut f\"ur Gravitationsphysik (Albert-Einstein-Institut) Am M\"uhlenberg 1, DE-14476 Potsdam, Germany
\label{addr5}
\and
Erasmus Mundus Joint Doctorate IRAP Ph.D. Student\label{addr6}
}

\date{Received: date / Accepted: date}

\maketitle

\begin{abstract}
We investigate the dynamics of test particles undergoing friction forces 
in a Friedmann-Robertson-Walker (FRW) spacetime.
The interaction with the background fluid is modeled by introducing a
Poynting-Robertson-like friction force in the equations of motion, leading to 
measurable (at least in principle) deviations of the particle trajectories 
from geodesic motion. 
The effect on the peculiar velocities of the particles is investigated for 
various equations of state of the background fluid and different standard cosmological models.
The friction force is found to have major effects on particle motion
in closed FRW universes, where it turns the time-asymptotic value (approaching the recollapse) of the
peculiar particle velocity from ultra-relativistic (close to light speed)
to a co-moving one, i.e., zero peculiar speed.
On the other hand, for open or flat universes the effect of the friction is not so significant, because
the time-asymptotic peculiar particle speed is largely non-relativistic also in the
geodesic case. 


\end{abstract}

\section{Introduction}

The search of a correct model describing our universe in the framework of general relativity is a central issue of modern cosmology, and a subject of constant investigation and debate (see, e.g., Refs. \cite{perlmutter,riess,peebles,ellis} and references therein). 
Similarly, the description of the dynamics of cosmological objects, like galaxies or galaxy clusters, in time-dependent gravitational fields strongly depends on either the particular cosmological 
model adopted or on possible acceleration mechanisms modifying the geodesic motion.
For all of these issues, a rich literature is available 
(see, e.g., Refs. \cite{mash1,mash2} and references therein).  

In this paper, we investigate the dynamics of massive test particles, undergoing friction effects in Fried\-mann-Robertson-Walker (FRW) spacetimes.
This is motivated by the possibility that non-gravitational interactions, such as 
collisions with the background fluid component, may be relevant in the description of the motion.
The fluid source of the FRW spacetime, modeling either an open, flat, or closed universe, is assumed to be described by a general equation of state, with pressure proportional to density through 
a constant parameter, $w$, representing baryonic matter, radiation, dark energy, etc., depending
on different choices of $w$.
The model equations governing particle dynamics are given in full generality.
As illustrative examples, for their numerical integration, we consider the case of pressureless matter, stiff matter and radiation. 

To the best of our knowledge, while applications to non-geodesic motion in black hole 
physics date back to the pioneering works of Poynting and Robertson \cite{poy1,poy2}, such  
an approach, involving viscous forces causing acceleration/deceleration of particles, has
not been adopted yet in a cosmological context. 
Applications to the motion of massive test particles in a Schwarzschild spacetime surrounded by either a perfect fluid or a thermal photon gas have been discussed in Refs. \cite{NOI1,NOI2}, respectively.
Particle dynamics in the Tolman metric generated by a photon gas source in thermodynamical equilibrium has also been investigated in Ref. \cite{NOI3}.

We study the effect of the friction on particle peculiar velocities, using an expression 
for the force term which is formally the same as the one introduced by Poynting and Robertson, 
and recently identified as a general relativistic extension of the Stokes law 
for classical viscous fluids \cite{NOI4}.  
Our analysis shows that deviations from geodesic motion can be very significant,
hence, in principle, also measurable, at least for the case of closed FRW universes.

\section{Basic equations of the model}

Let us consider the Friedmann metric\footnote{Here greek indices run from $0$ to $3$ whereas latin ones from $1$ to $3$. We also use geometrized units with $c=G=\hbar=1$. The signature of the metric is $[-,+,+,+]$.}  written in Robert\-son-Walker comoving coordinates \cite{exact}
\beq
ds^2=-dt^2+a^2\left[dr^2+\Sigma^2(d\theta^2+\sin^2\theta d\phi^2) \right]\,, 
\eeq
where $a=a(t)$ is the scale factor and $\Sigma=[\sin r, r, \sinh r]$ corresponding to closed, flat and open universes, respectively.
The matter content of the universe is described by the perfect fluid stress-energy tensor
\beq
\label{stress-energy}
T_{\alpha\beta}=(\rho+p)u_\alpha u_\beta+pg_{\alpha\beta}\,, \qquad u_\alpha u^\alpha=-1\,,
\eeq
where $u=\partial_t$ and the pressure $p=p(t)$ and the energy density $\rho=\rho(t)$ do not depend on the spatial coordinates, but only on the time, 
due to the homogeneity and isotropy of the space. 
Einstein's field equations $G_{\mu\nu}=8\pi T_{\mu\nu}$ can be summarized by
the energy conservation equation 
\beq
\label{energy}
\dot\rho=-3\frac{\dot a}{a}(\rho +p)\,,
\eeq
and the Friedmann equation
\beq
\label{friedeq}
\dot a^2  = -k +\frac83 \pi \rho  a^2\,,
\eeq
where $k=-(\Sigma'{}^2-1)/\Sigma^2=-\Sigma ''/\Sigma=[1,0,-1]$ for the case of closed, flat and open universes, respectively.
A dot here denotes derivative with respect to time, while a prime derivative with respect to $r$.

Hereafter, the source fluid (single fluid for simplicity) is set to obey an equation of state of the form
\beq
p=w\rho\,, \qquad
w=const.\,,
\eeq
since this equation is general enough to discuss several interesting physical situations. 
In fact, the case $w=1$ mimics a stiff matter, $w=1/3$ a radiation field, $w=0$ a massive dust, $w\in [-1,-1/3]$ a quintessence field, $w=-1$ a cosmological constant term and finally $w<-1$ a phantom energy field. 
Eq. (\ref{energy}) then gives
\beq 
\label{energy2}
\rho = \rho_0 \left(\frac{a_0}{a}\right)^{3(1+w)}\,,
\eeq
where $\rho_0$ and $a_0$ are the energy density and the scale factor today.

It is convenient to introduce an orthonormal frame adapted to the comoving observers (i.e., a comoving frame), whose world lines are the coordinate time lines
\begin{eqnarray}
e_{\hat t}=\partial_t,\,\,\,
e_{\hat r}= \frac {1}{a}\partial_r,\,\,\,
e_{\hat \theta}=\frac {1}{a\Sigma}\partial_\theta,\,\,\,
e_{\hat \phi}=\frac {1}{a\Sigma\sin\theta}\partial_\phi\,.
\end{eqnarray}
The $4$-velocity of a test particle is given by 
\beq
U=\gamma \left(e_{\hat t}+\nu^{\hat a}e_{\hat a}\right)\,, \qquad \gamma=\frac{1}{\sqrt {1-\nu^2}}\,,
\eeq
where $\vec\nu=\nu^{\hat a}e_{\hat a}$ is the spatial ``peculiar velocity'' vector of the particle relative to the family of comoving observers with magnitude $\nu=\sqrt{\delta_{\hat a\hat b}\nu^{\hat a}\nu^{\hat b}}$ and $\gamma$ is the corresponding Lorentz factor.
Let $P=mU$ be the associated peculiar 4-momentum, $m$ denoting the particle mass.
As it is well known, in the FRW models the frame components of the spatial momentum behave as $a(t)^{-1}$, expressing the law of decay of peculiar velocities as the universe expands (see, e.g., Ref. \cite{peebles}).

We will examine the case in which $U$ is accelerated (or decelerated), due to the interaction of the particles with the background fluid, by the following friction force (treated as an additional external force of Poynting-Robertson-like form \cite{poy1,poy2})
\beq
\label{f_scatt_def}
f_{\rm (fric)}(U)^\alpha =-\sigma P(U)^{\alpha}{}_{\mu}T^{\mu \nu}U_{\nu}\,,
\eeq
where $\sigma$ is the cross section of the process, $P(U)^{ \alpha}{}_{ \nu}=\delta^{ \alpha}{}_{ \nu}+U^{ \alpha} U_{ \nu}$  projects orthogonally to $U$ and $T^{\mu \nu}$ is given by equation (\ref{stress-energy}). 
Explicitly, we have
\begin{eqnarray}
\label{ffric}
f_{\rm (fric)}(U)&=& 
-\sigma\gamma^3 (\rho+p)(\nu^2 e_{\hat t}+\nu^{\hat a} e_{\hat a})\nonumber\\
&=&-\sigma(1+w)\gamma^2 \rho\nu \bar U\,,
\end{eqnarray}
where $\bar U=\gamma (\nu e_{\hat t}+\frac{1}{\nu}\nu^{\hat a} e_{\hat a})$ is a unit spacelike vector orthogonal to $U$ in the plane of the motion. 

Recently, this force has been identified with the relativistic generalization of the Stokes force acting on a body moving in a viscous fluid \cite{NOI4}. 
Interestingly, this force switches its sign when $w=-1$ (i.e., at the phantom energy boundary value): for $w<-1$ (even if one may consider this as an \lq\lq exotic" case) the friction force reverses its sign, so that the test particle extracts energy from the cosmic fluid.
In other words, the cosmic fluid becomes an {\it active} media \cite{ACTIVE}.

Denoting by $a(U)=\nabla_{U}U$ the $4$-acceleration of a test particle, the equations 
of motion read as follows 
\beq
ma(U)^{\alpha}=f_{\rm (fric)}(U)^\alpha\,.  
\eeq
Explicitly, we have
\begin{eqnarray}
\label{eqmotogen}
\frac {d\nu^{\hat r}}{d\tau}&=& \frac{\gamma }{a \Sigma}\left[((\nu^{\hat \theta})^2+(\nu^{\hat \phi})^2)\Sigma'-\frac{\nu^{\hat r} \Sigma \dot a}{\gamma^2} \right] -A\nu^{\hat r}(\rho+p)\,, \nonumber \\
\frac {d\nu^{\hat \theta}}{d\tau}&=&\frac{\gamma  }{a \Sigma}\left[-\nu^{\hat \theta}\nu^{\hat r}\Sigma'-\frac{\nu^{\hat \theta} \Sigma \dot a}{\gamma^2}+\cot\theta (\nu^{\hat \phi})^2 \right] \nonumber\\
&&-A\nu^{\hat \theta}(\rho+p)\,, \\
\frac {d\nu^{\hat \phi}}{d\tau}&=& -\frac{\gamma \nu^{\hat \phi} }{a \Sigma}\left[\nu^{\hat r}\Sigma'+ \frac{ \Sigma \dot a}{\gamma^2}+\cot\theta \nu^{\hat \theta} \right] -A\nu^{\hat \phi}(\rho+p)\,, \nonumber
\end{eqnarray}
which must be completed by the evolution equations
\begin{eqnarray}
\label{evoleqs}
\frac {dt}{d\tau}&= \gamma\,, &\quad
\frac {dr}{d\tau}= \frac{\gamma\nu^{\hat r} }{a}\,,  \nonumber\\ 
\frac {d\theta}{d\tau}&=\displaystyle\frac{\gamma}{a\Sigma}\nu^{\hat \theta}\,, &\quad   
\frac {d\phi}{d\tau}=\frac{\gamma }{a\Sigma\sin\theta }\nu^{\hat \phi}\,.
\end{eqnarray}
The coupling constant between the particle and the field is given by 
\beq
A=8\pi\frac {\sigma}{m}\,.
\eeq
The spherical symmetry of the problem allows us to restrict our analysis to a planar motion also in presence of interactions, so that we set $\theta=\pi/2$ and $\nu^{\hat \theta}=0$. 
Moving to a polar representation of the velocity
\beq
\nu^{\hat r}= \nu\sin\alpha\,,\qquad \nu^{\hat \phi}= \nu\cos\alpha\,,
\eeq
the equations of motion (\ref{eqmotogen}) and the evolution equations (\ref{evoleqs}) become
\begin{eqnarray}
&&\frac {d\nu}{d\tau}= -\nu \left[ A(1+w)\rho+\frac{\dot a}{a\gamma}\right]\,,\quad 
\frac {d\alpha}{d\tau}= \frac{\gamma \nu\Sigma' \cos\alpha  }{a \Sigma}\,, \nonumber\\
\label{eqalpha}
&&\frac {dr}{d\tau}=\frac {\gamma\nu \sin\alpha  }{a}\,, \quad 
\frac {d\phi}{d\tau}=\frac {\gamma }{ a\Sigma}\nu \cos \alpha \,.
\end{eqnarray}
Inspection of these equations shows that the dependence on the friction parameter $A$ only affects the equation for $\nu$.

Expressing the derivatives with respect to the proper time in terms of those with respect to the coordinate time we finally obtain 
\begin{eqnarray}
\label{finale}
&&\dot \nu=-\frac{\nu}{\gamma}\left[A(1+w)\rho+\frac{1}{\gamma}\frac{\dot a}{a}\right]\,,\nonumber\\
&&\dot \alpha= \frac{\nu\Sigma' \cos\alpha  }{a \Sigma} \,, \quad 
\dot r=\frac {\nu \sin\alpha  }{a}\,, \quad 
\dot \phi=\frac {\nu \cos \alpha }{ a\Sigma} \,, \nonumber\\
&&\dot a=\sqrt{ -k +\frac83 \pi \rho  a^2}\,,
\end{eqnarray}
where the energy density is a function of the scale factor, as from Eq. (\ref{energy2}).
We can also identify in the first equation above the contributions of the spatial acceleration due to friction and gravitation, namely
\beq
\dot \nu=a_{\rm(fric)} +a_{\rm(grav)}\,, 
\eeq
with
\beq
a_{\rm(fric)}=-A(1+w)\rho\frac{\nu}{\gamma}\,, \qquad
a_{\rm(grav)}=-\frac{\nu}{\gamma^2}\frac{\dot a}{a}\,.
\eeq
Studying the behavior of their ratio as a function of coordinate time allows to determine the dominance of one term on the other during the evolution.
Moreover, in the same equation one easily recognizes the two equilibrium solutions $\nu=0$ and $\nu=1$, corresponding to the limiting cases of comoving matter and ultrarelativistic matter, respectively.
These equilibrium solutions already exist in the geodesic case and only the way in which they are approached is modified by the presence of the friction force.
For a closed universe, $\dot a$ changes its sign (passing from positive to negative values) as soon as the universe starts to recollapse.
As a result, the gravitational acceleration term $a_{\rm(grav)}$ contributes to a positive increasing of $\dot\nu$.
Therefore, the fate of geodesics is to end necessarily their evolution at $\nu=1$ (see Fig. \ref{fig:1}).
The friction acceleration $a_{\rm(fric)}$, instead, only gives a monotonically negative contribution to $\dot\nu$, which is but the dominant behavior at times close to the Big Crunch where $a\to0$, i.e., $\rho\to\infty$ for ordinary matter (implying $\nu\to0$, see the behavior of non-geodesic curves in Fig. \ref{fig:1}).
On the contrary, for both flat and open universes, $\dot a$ is always positive, leading to $\nu\to0$ asymptotically (see Fig. \ref{fig:2}). 

System (\ref{finale}) admits the integral of motion
\beq
\label{vincalpha}
\Sigma\cos\alpha=\Sigma_0\cos\alpha_0\,,
\eeq
as it readily follows from the equation
\beq
\frac{d\alpha}{dr}=\cot \alpha \frac{\Sigma '}{\Sigma}\,.
\eeq 
Eq. (\ref{vincalpha}) must hold for every value of $r$ and $\alpha$, in particular for $r=0$, i.e., $\Sigma=0$.
Therefore, it implies $\alpha=\pm\pi/2$ for all times, so that Eqs. (\ref{finale}) also yield
\beq
\label{finaler}
\dot r=\pm\frac {\nu}{a}\,,\qquad
\phi=\phi_0\,.
\eeq
This is a consequence of the spatial isotropy and homogeneity of the FRW universe, which imply that the geodesics issuing from any point in space taken as spatial origin of spherical polar coordinates are purely radial.
Furthermore, one can choose the plus sign in Eq. (\ref{finaler}), corresponding to the increasing with time of the radial parameter.

Let us consider first the case of geodesic motion ($A=0$).
In the case of comoving matter we simply have $\nu=0$, because $u=e_{\hat t}=\partial_t$ is a geodesic 4-vector, implying that $r=$ const.
The proper distance from the origin to a galaxy at matter-comoving radial coordinate $r$ at time $t$ is given by $D=ar$, so that its \lq\lq coordinate" velocity away from us is given by Hubble's law $v\equiv \dot D=\dot a r=HD$, where $H=\dot a/a$ is the Hubble scalar.
In the case of non-comoving matter (i.e., $\nu\not = 0$), instead, $r$ also depends on time according to Eq. (\ref{finaler}), so that Hubble's law is modified as $v=HD+\nu$.
Galaxies are thus not expected to follow Hubble's flow exactly in this case, but only to approach it at late times during the evolution \cite{peebles}.
In fact, in addition to the expansion of the universe, galaxy motions are affected by the gravity of specific, nearby structures.
For instance, for galaxies in the Local Group, stars inside the Milky Way, and for objects in the Solar System their peculiar velocity is larger than their expansion velocity, implying that the peculiar term dominates the total coordinate velocity. 
For distant galaxies, instead, peculiar velocities have in general very small values, so that Hubble's law applies in this case.
Peculiar velocities are also expected to be generated by some inhomogeneities in the matter distribution.
Spatial inhomogeneities can in turn significantly modify the structure of cosmic jets as discussed in Refs. \cite{mash1,mash2}, with particles being accelerated to very high peculiar velocities.

The first of Eqs. (\ref{finale}), giving the magnitude of the particle peculiar velocity, is readily integrated 
\beq
\label{nugeosol}
\nu=\frac{\nu_0}{\sqrt{\nu_0^2+(a/a_0)^2(1-\nu_0^2)}}\,.
\eeq
For a closed universe, $a\to0$ at the Big Crunch, and therefore $\nu\to1$.
For both flat and open universes, instead, $a\to\infty$ asymptotically, implying $\nu\to0$.
The only nonvanishing frame component of the spatial peculiar 4-momentum turns out to be given by
\beq
P^{\hat r}=\gamma\nu=\gamma_0\nu_0\frac{a_0}{a}\,,
\eeq
which is just the law of variation of peculiar velocities in standard cosmological models, as stated before.
Substituting then into Eq. (\ref{finaler}), a quadrature delivers the following solution for $r$:
$$
r=\frac{\nu_0}{a_0H_0}\int_0^{\frac{a}{a_0}}\frac{dx}{x\sqrt{\left(\nu_0^2+\displaystyle\frac{x^2}{\gamma_0^2}\right)\left(1-\Omega_0+\displaystyle\frac{\Omega_0}{x^{1+3w}}\right)}}\,,
$$
where $\Omega_0=8\pi\rho_0/(3H_0^2)$ is the value of the density parameter 
at the current epoch $t=t_0$ and $H_0$ is the Hubble constant. 
The remaining observational parameters are given by the present values of the curvature 
density $\Omega_{k\,0}$ and of the deceleration parameter $q_0$
\beq
\Omega_{k\,0}=-\frac{k}{a_0^2H_0^2}=1-\Omega_0\,,\qquad
q_0=\frac12\Omega_0(1+3w)\,,
\eeq
so that $\Omega_0>1$, $\Omega_0=1$ and $\Omega_0<1$ correspond to closed, flat and open universes, respectively.

We will investigate in the following how the interaction of the particles with the background fluid modifies the behavior of the peculiar velocity (and thus Hubble's law) as a function of time with respect to the case of non-comoving geodesic matter. 
We are not interested in discussing the more general case of a universe consisting of a mixture of different fluids corresponding to the different epochs, but only in studying the effect of the friction force in simple situations; the extension will be straightforward.
Therefore, we will consider the case of a single fluid component in each kind of FRW universe separately.
Note that the choice of initial conditions, as well as observational parameters adopted in the numerical integrations below, is suitable to highlight the deviation from the case of geodesic motion.

\subsection{{\it Closed universe}}

Let us consider first the Friedmann equation (\ref{friedeq}) with energy density (\ref{energy2}).
Through the transformation (see, e.g., Ref. \cite{kolbturner})
\beq
\label{Thdef}
1-\cos\Theta=\frac{2}{\Omega_0}(\Omega_0-1)\left(\frac{a}{a_0}\right)^{1+3w}\,,
\eeq
with $q_0>(1+3w)/2$, $w>-1/3$ (the case $w<-1/3$ can be treated similarly), it can be cast in the form
\beq
\label{eqTh}
\dot\Theta=\frac{H_0J_{\Theta}}{(1-\cos\Theta)^{1/(1+3w)}}\,,
\eeq
with
\beq
J_{\Theta}=(1+3w)\sqrt{\frac{\Omega_0}{2}}\left[\frac{2}{\Omega_0}(\Omega_0-1)\right]^{\frac{3(1+w)}{2(1+3w)}}\,.
\eeq
The latter equation can be integrated for selected values of $w$, providing 
the solution for the scale factor as a function of time in parametric form.

Consider then the equations of motion.
Looking for $\nu$ and $r$ as functions of $\Theta$ implies
\begin{eqnarray}
\label{closedfin}
\frac {d\nu}{d\Theta}&=&-\frac{\nu\sqrt{1-\nu^2}}{(1+3w)(1-\cos\Theta)}\left[\sin\Theta\sqrt{1-\nu^2}\right.\nonumber\\
&&\left.
+{\tilde A}\left(\frac{1+w}{1+3w}\right)\frac{J_{\Theta}}{(1-\cos\Theta)^{1/(1+3w)}}\right]\,,\nonumber\\
\frac {dr}{d\Theta}&=&\frac{\nu}{1+3w}\,,
\end{eqnarray}
with ${\tilde A}={3AH_0}/{4\pi}$ dimensionless, which can be numerically integrated.
Note that from Eq. (\ref{Thdef}) we have $a=0$, i.e., $\Theta=0$, for $t=0$, whereas $a=a_0$, i.e., $\cos\Theta=(1+3w-q_0)/q_0$, for $t=t_0$.
Finally, the explicit dependence on time is given by Eq. (\ref{eqTh}), which can be written 
in dimensionless form by introducing the rescaling $\tilde t=H_0t$ as follows
\beq
\label{eqTh2}
\frac{d\tilde t}{d\Theta}=\frac1{J_{\Theta}}(1-\cos\Theta)^{1/(1+3w)}\,.
\eeq

\subsection{{\it Open universe}}

By using the transformation (see, e.g., Ref. \cite{kolbturner})
\beq
\label{Psidef}
\cosh\Psi-1=\frac{2}{\Omega_0}(1-\Omega_0)\left(\frac{a}{a_0}\right)^{1+3w}\,,
\eeq
with $0\le q_0<(1+3w)/2$, $w>-1/3$, the Friedmann equation can be cast in the form
\beq
\label{eqPsi}
\dot\Psi=\frac{H_0J_{\Psi}}{(\cosh\Psi-1)^{1/(1+3w)}}\,,
\eeq
with
\beq
J_{\Psi}=(1+3w)\sqrt{\frac{\Omega_0}{2}}\left[\frac{2}{\Omega_0}(1-\Omega_0)\right]^{\frac{3(1+w)}{2(1+3w)}}\,.
\eeq
The latter equation can be integrated for selected values of $w$, giving the solution 
for the scale factor as a function of time in parametric form.

Consider then the equations of motion.
Looking for $\nu$ and $r$ as functions of $\Psi$, implies
\begin{eqnarray}
\label{openfin}
\frac {d\nu}{d\Psi}&=&-\frac{\nu\sqrt{1-\nu^2}}{(1+3w)(\cosh\Psi-1)}\left[\sinh\Psi\sqrt{1-\nu^2}\right.\nonumber\\
&&\left.
+{\tilde A}\left(\frac{1+w}{1+3w}\right)\frac{J_{\Psi}}{(\cosh\Psi-1)^{1/(1+3w)}}\right]\,,\nonumber\\
\frac {dr}{d\Psi}&=&\frac{\nu}{1+3w}\,,
\end{eqnarray}
which can be numerically integrated.
Note that from Eq. (\ref{Psidef}) we have $a=0$, i.e., $\Psi=0$, for $t=0$, whereas $a=a_0$, i.e., $\cosh\Psi=(1+3w-q_0)/q_0$, for $t=t_0$.
Finally, the explicit dependence on time is given by Eq. (\ref{eqPsi}), which can be rewritten as
\beq
\label{eqPsi2}
\frac{d\tilde t}{d\Psi}=\frac1{J_{\Psi}}(\cosh\Psi-1)^{1/(1+3w)}\,.
\eeq

\subsection{{\it Flat universe}}

The solution for the scale factor is given by
\beq
a(t)=a_0\left(\frac{t}{t_0}\right)^{\frac{2}{3(1+w)}}\,, \quad
t_0=\frac{2}{3(1+w)H_0}\,,
\eeq
whereas the energy density turns out to be
\beq
\rho(t)=\rho_0\left(\frac{t_0}{t}\right)^2\,, \qquad
\rho_0=\frac{3H_0^2}{8\pi}\,.
\eeq
The equations of motion then become
\begin{eqnarray}
\label{finaleflat}
\frac {d\nu}{d\tilde t}&=&
-\frac23\frac{\nu\sqrt{1-\nu^2}}{(1+w)\tilde t}\left[\frac{\tilde A}{3\tilde t}+\sqrt{1-\nu^2}\right]\,,\nonumber\\
\frac {dr}{d\tilde t}&=&
\nu\left[\frac{2}{3(1+w)\tilde t}\right]^{\frac{2}{3(1+w)}}\,.
\end{eqnarray}

\section{Numerical integration for different $w$}

Let us analyze how the different kind of energy content of the universe affects the friction force and then the motion of test particles.
In particular, the cases of a matter-dominated ($w=0$ and $w=1$, for instance) and radiation-dominated universe ($w=1/3$) will be considered.
The case of a cosmological constant ($\Lambda$-dominated universe, $w=-1$) is not relevant for the present analysis, since the friction force identically vanishes, as from Eq. (\ref{ffric}), so that particles move along geodesics.

\subsection{{\it Closed universe}}

The equations of motion to be numerically integrated are given by Eqs. (\ref{closedfin})--(\ref{eqTh2}).
The latter provides the relation between the parameters $\Theta$ and $\tilde t$. 
The value $\Theta=0$ corresponds to $\tilde t=0$, whereas the value $\Theta=2\pi$ to the time of recollapsing of the universe, which is different depending on the selected values of $w$ and $q_0$.
For instance, for $w=0$, Eq. (\ref{eqTh2}) gives 
\beq
\tilde t=\frac{\Omega_0}{2}(\Omega_0-1)^{-3/2}(\Theta-\sin\Theta)\,.
\eeq
The value of $\tilde t$ at the Big Crunch is then 
\beq
\tilde t_{BC}=\pi\Omega_0(\Omega_0-1)^{-3/2}\,.
\eeq
For $w=1/3$ we have instead 
\beq
\tilde t=\frac{\sqrt{\Omega_0}}{\Omega_0-1}\left(1-\cos\frac{\Theta}{2}\right)\,,
\eeq
with
\beq
\tilde t_{BC}=\frac{2\sqrt{\Omega_0}}{\Omega_0-1}\,.
\eeq

\begin{figure*} 
\typeout{*** EPS figure 1}
\begin{center}
$\begin{array}{cc}
\includegraphics[scale=0.3]{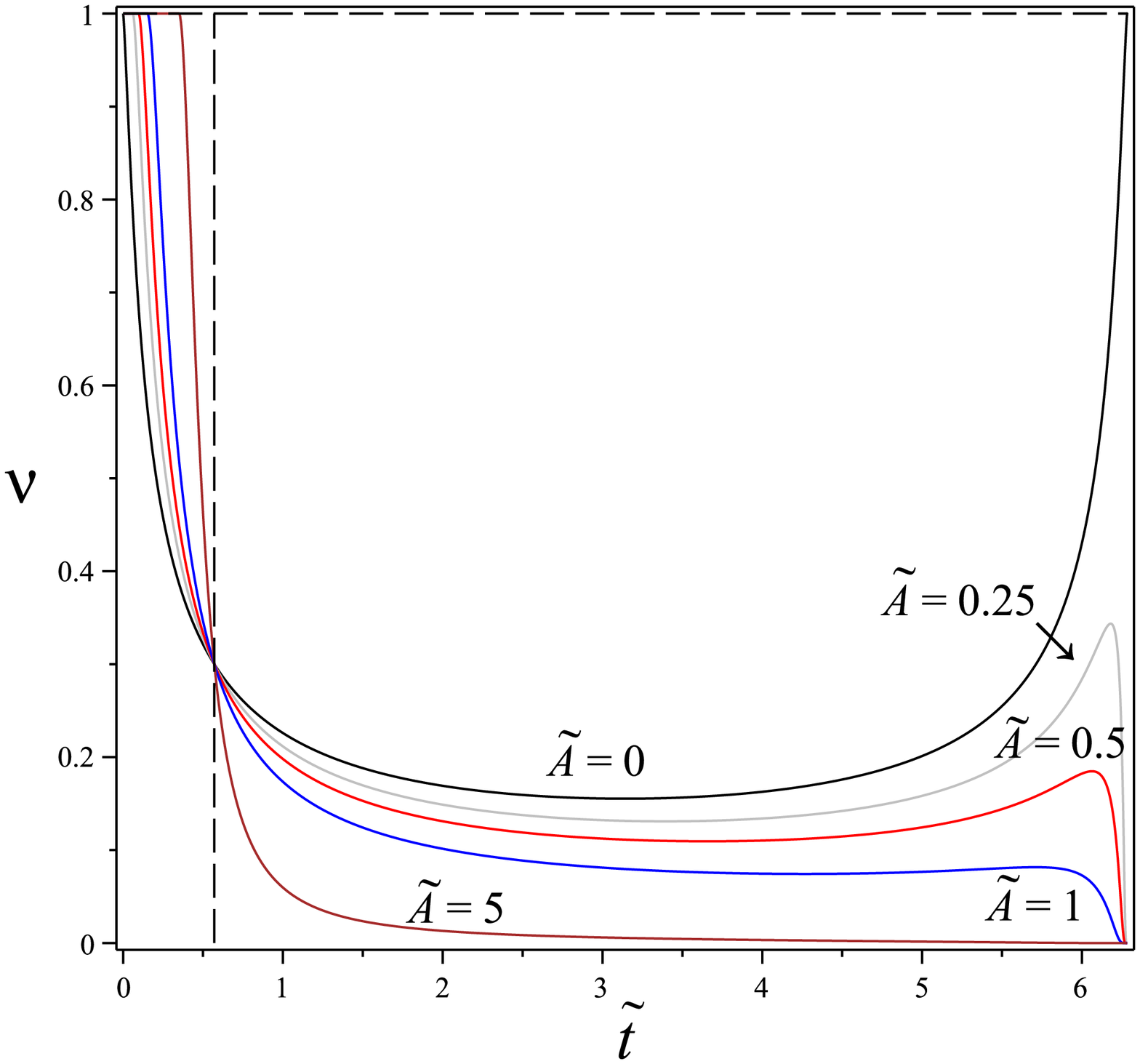}&\quad
\includegraphics[scale=0.3]{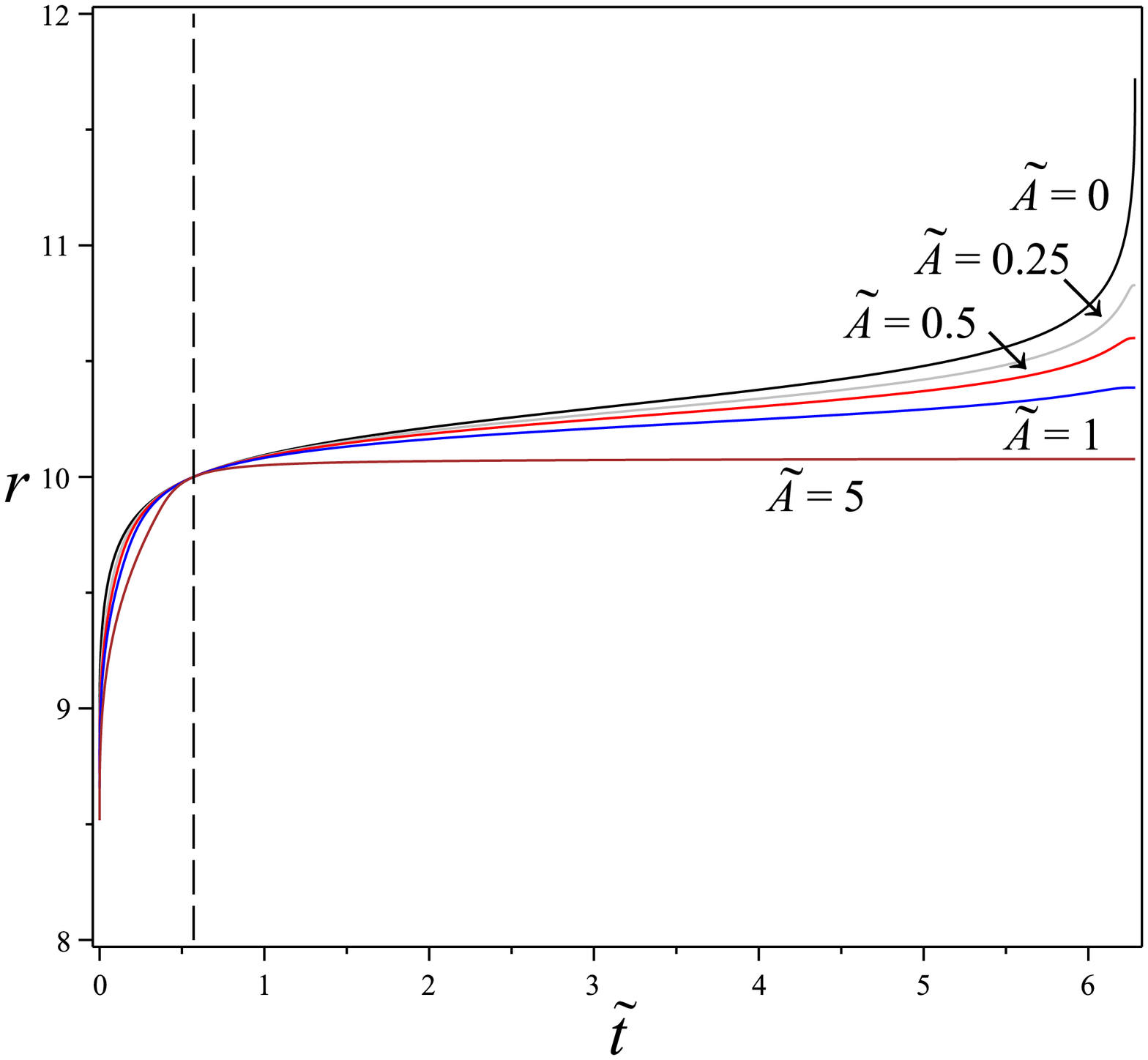}\\
\quad\mbox{(a)}\quad[w=0]\quad &\quad \mbox{(b)}\quad[w=0]\\[.4cm]
\includegraphics[scale=0.3]{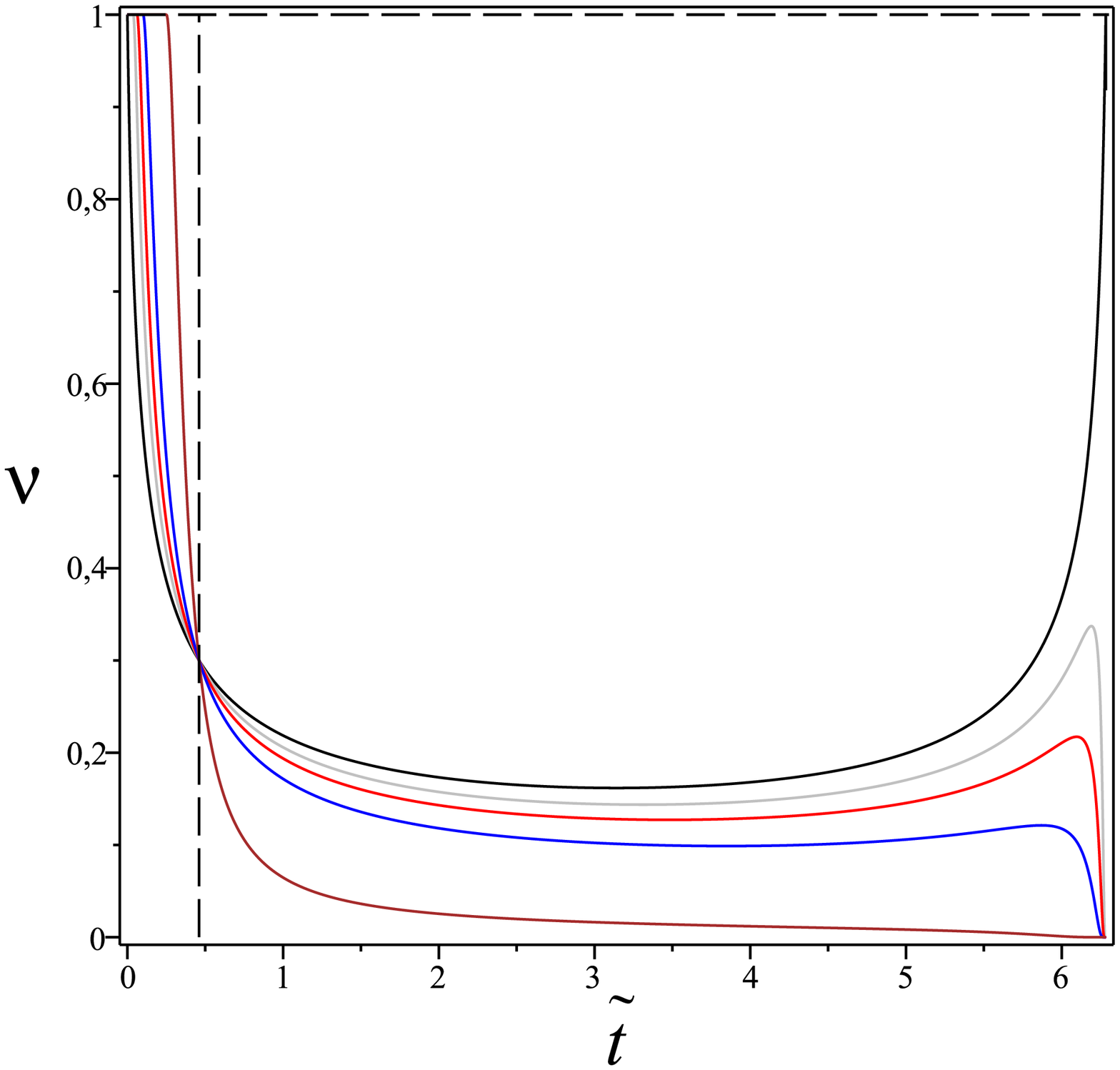}&\quad
\includegraphics[scale=0.3]{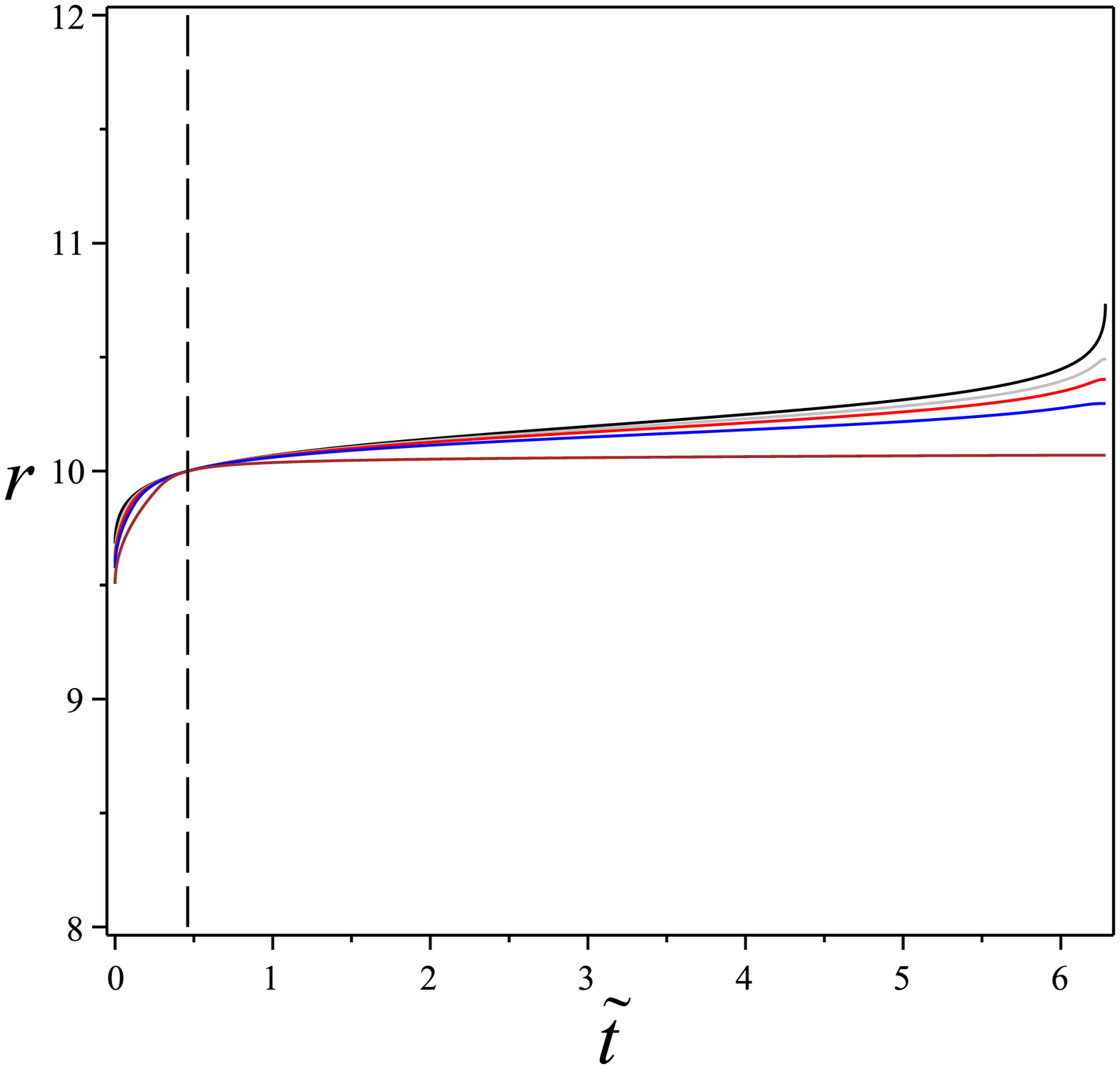}\\
\quad\mbox{(c)}\quad[w=1/3]\quad &\quad \mbox{(d)}\quad[w=1/3]\\[.4cm]
\includegraphics[scale=0.3]{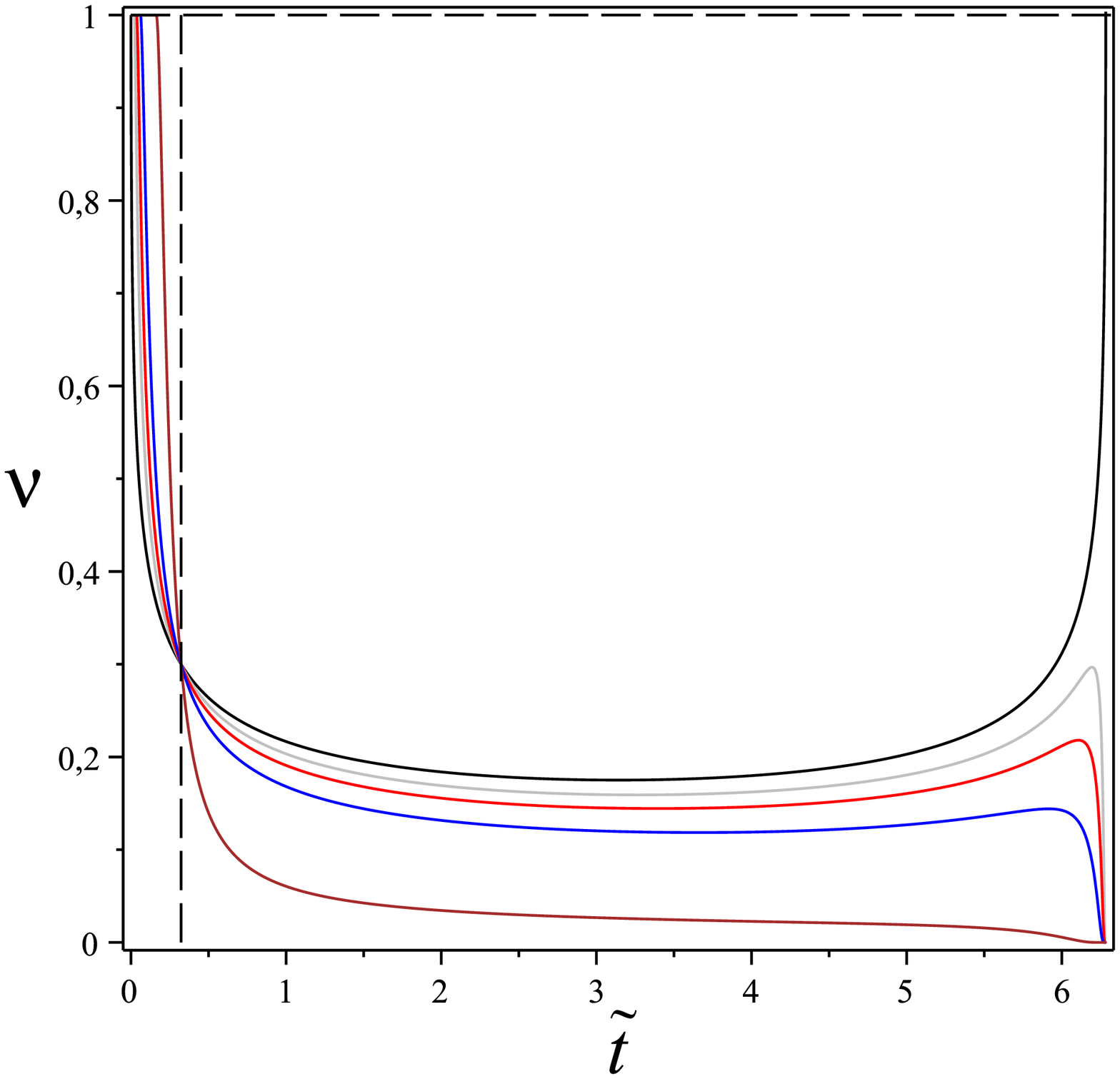}&\quad
\includegraphics[scale=0.3]{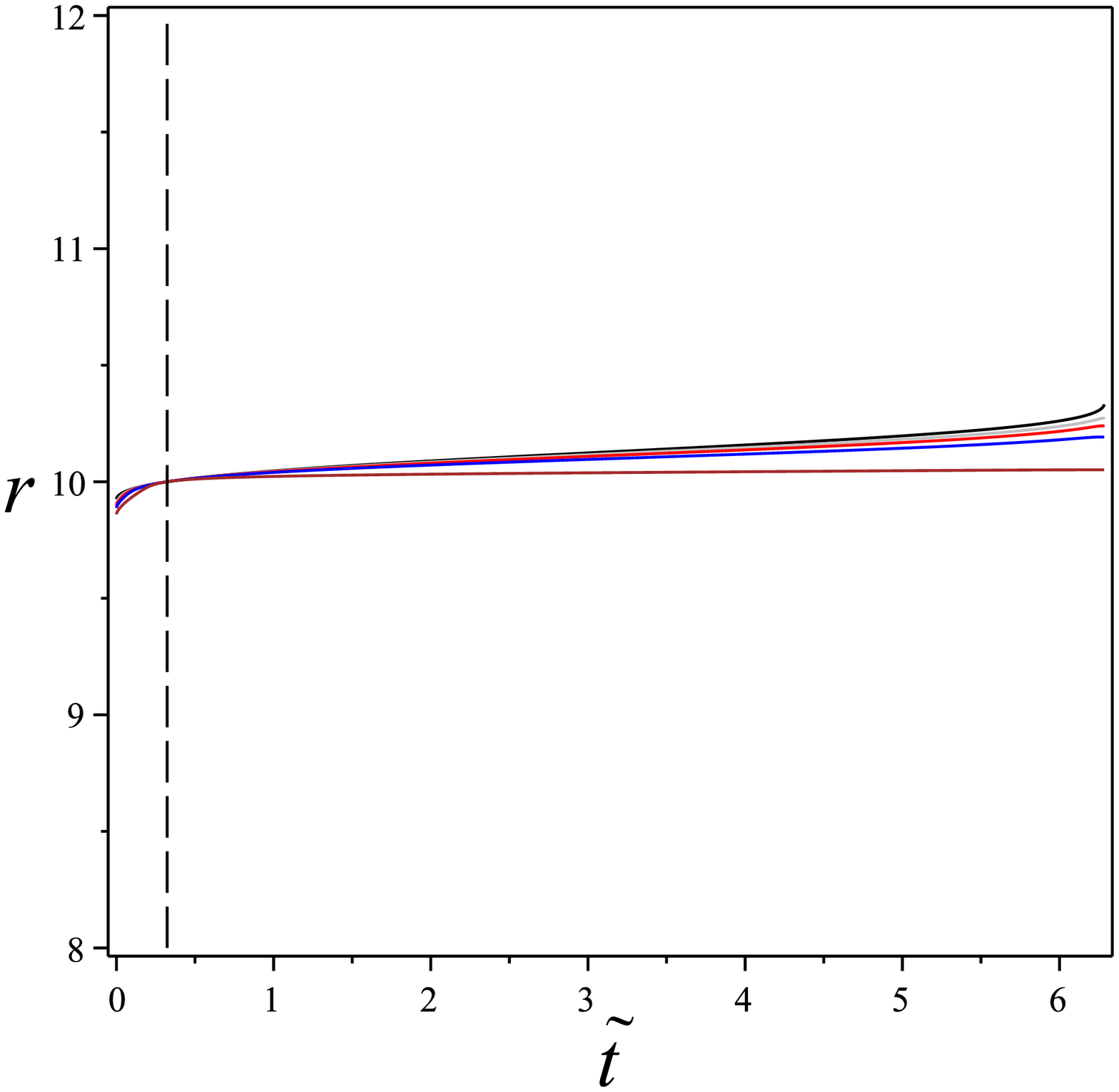}\\
\quad\mbox{(e)}\quad[w=1]\quad &\quad \mbox{(f)}\quad[w=1]\\
\end{array}$
\end{center}
\caption{[Closed universe]
The behaviors of the linear velocity $\nu$ and the coordinate $r$ as functions of the time parameter $\tilde t=H_0t$ are shown for a closed FRW universe for selected values of $w=[0,1/3,1]$ and different values of the friction parameter $\tilde A=[0,0.25,0.5,1,5]$.
The set of equations (\ref{closedfin})--(\ref{eqTh2}) has been numerically integrated with initial conditions $\nu_0=\nu(\Theta_0)=0.3$, $r_0=r(\Theta_0)=10$ and $\tilde t_0=\tilde t(\Theta_0)$, the latter depending on the choice of $w$ and $q_0$.
Panels (a) and (b) refer to $w=0$ and $q_0=1$, so that $\Theta_0=\pi/2$ and $\tilde t_0\approx0.571$.
Panels (c) and (d) refer to $w=1/3$ and $q_0\approx1.373$, so that $\Theta_0\approx1.097$ and $\tilde t_0\approx0.460$.
Panels (e) and (f) refer to $w=1$ and $q_0\approx2.228$, so that $\Theta_0\approx0.651$ and $\tilde t_0\approx0.323$.
Different values of $\tilde A$ label the various curves of panels (c)--(f) as in panels (a) and (b).
The value of the deceleration parameter for each case has been chosen in such a way that the time parameter at the Big Crunch have the common value $\tilde t_{BC}=2\pi$ for a better comparison.
The vertical dashed line in every plot corresponds to the present time $\tilde t=\tilde t_0$. 
The curves in the region $\tilde t<\tilde t_0$ are the analytic extension in the past of the solutions.
The equilibrium solutions at $\nu=0$ and $\nu=1$ (marked with a horizontal dashed line) are future and past attractors, respectively.    
}
\label{fig:1}
\end{figure*}

The behaviors of $\nu$ and $r$ as functions of $\tilde t$ are shown in Fig. \ref{fig:1} for selected values of $w=[0,1/3,1]$ and different values of the friction parameter. 
The value of the deceleration parameter for each case has been chosen in such a way that the time parameter at the Big Crunch have the common value $\tilde t_{BC}=2\pi$ for a better comparison.
It turns out that the numerical integration of the equations of motion for $w=1/3$ and $w=1$ does not show any significant difference with respect to the case $w=0$.
The effect of the friction force dominates at early as well as late times.
As a consequence, from Fig. \ref{fig:1} we see a drastic change between geodesic and non-geodesic motion especially when the universe recollapses.
In fact, approaching that epoch, geodesics tend to become ultrarelativistic, whereas particles undergoing friction force tend to a rest state.
The onset of this behavior occurs earlier as the interaction strength $\tilde A$ increases. 
The deviation of the peculiar geodesic and non-geodesic velocities is less evident integrating backward in time, in which case both of them approach the ultrarelativistic behavior.


\subsection{{\it Open universe}}

The equations of motion to be numerically integrated are given by Eqs. (\ref{openfin})--(\ref{eqPsi2}).
The latter provides the relation between the parameters $\Psi$ and $\tilde t$.
For instance, for $w=0$, we obtain
\beq
\tilde t=\frac{\Omega_0}{2}(1-\Omega_0)^{-3/2}(\sinh\Psi-\Psi)\,,
\eeq
whereas for $w=1/3$
\beq
\tilde t=\frac{\sqrt{\Omega_0}}{1-\Omega_0}\left(\cosh\frac{\Psi}{2}-1\right)\,.
\eeq

\begin{figure*}
\begin{center}
$\begin{array}{cc}
\includegraphics[scale=0.3]{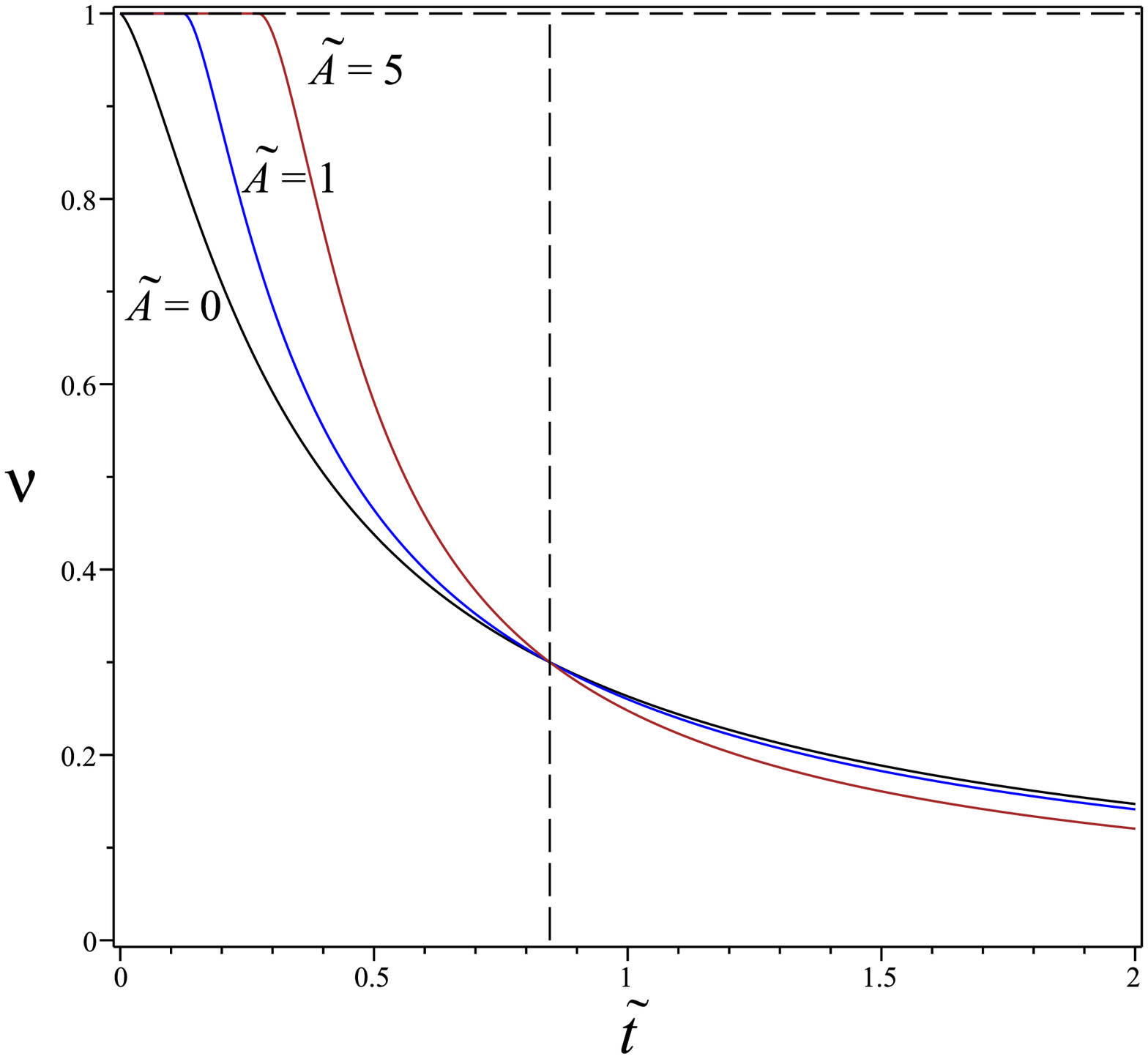}&\quad
\includegraphics[scale=0.3]{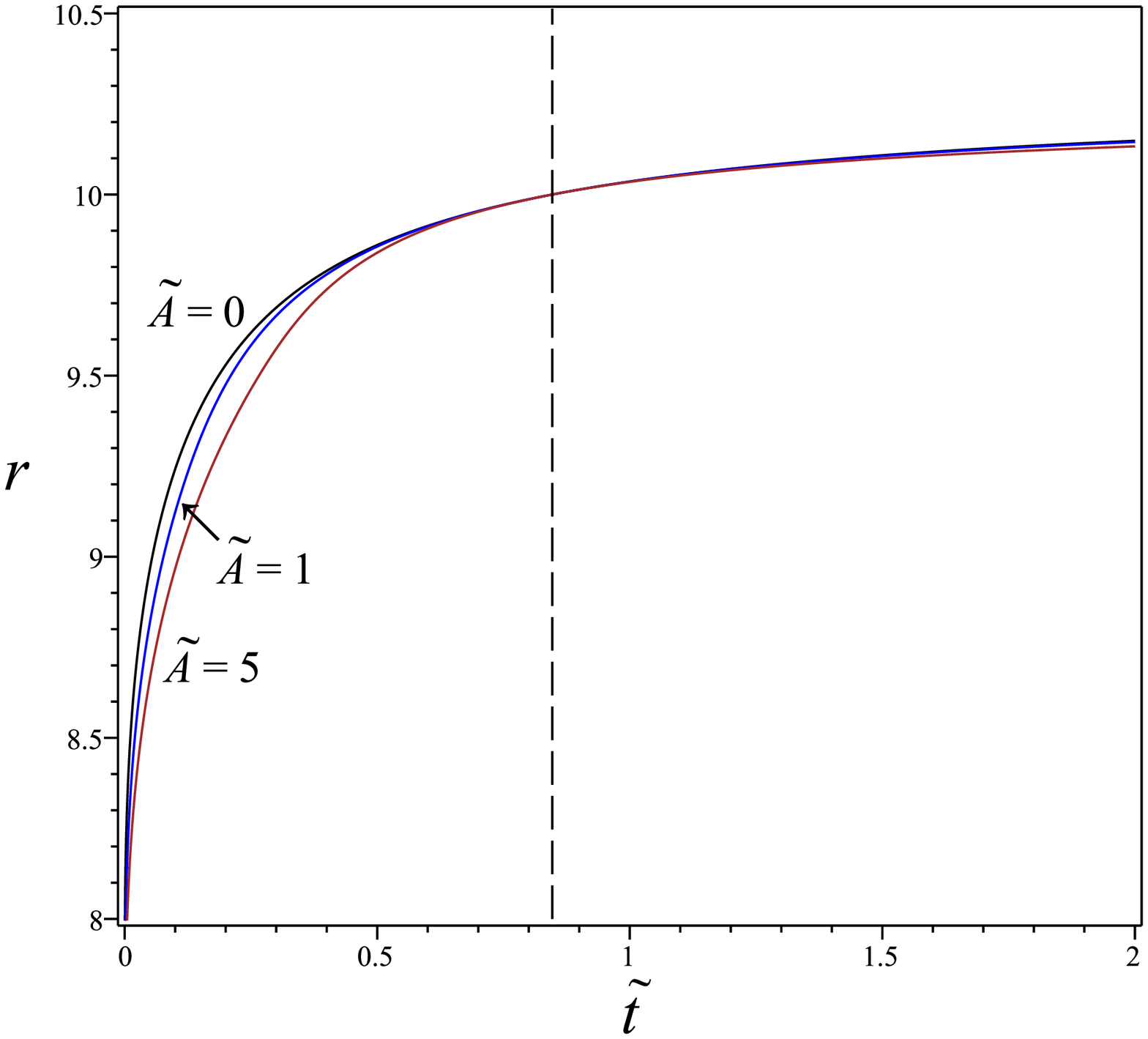}\\
\quad\mbox{(a)}\quad[w=0]\quad &\quad \mbox{(b)}\quad[w=0]\\[.4cm]
\includegraphics[scale=0.3]{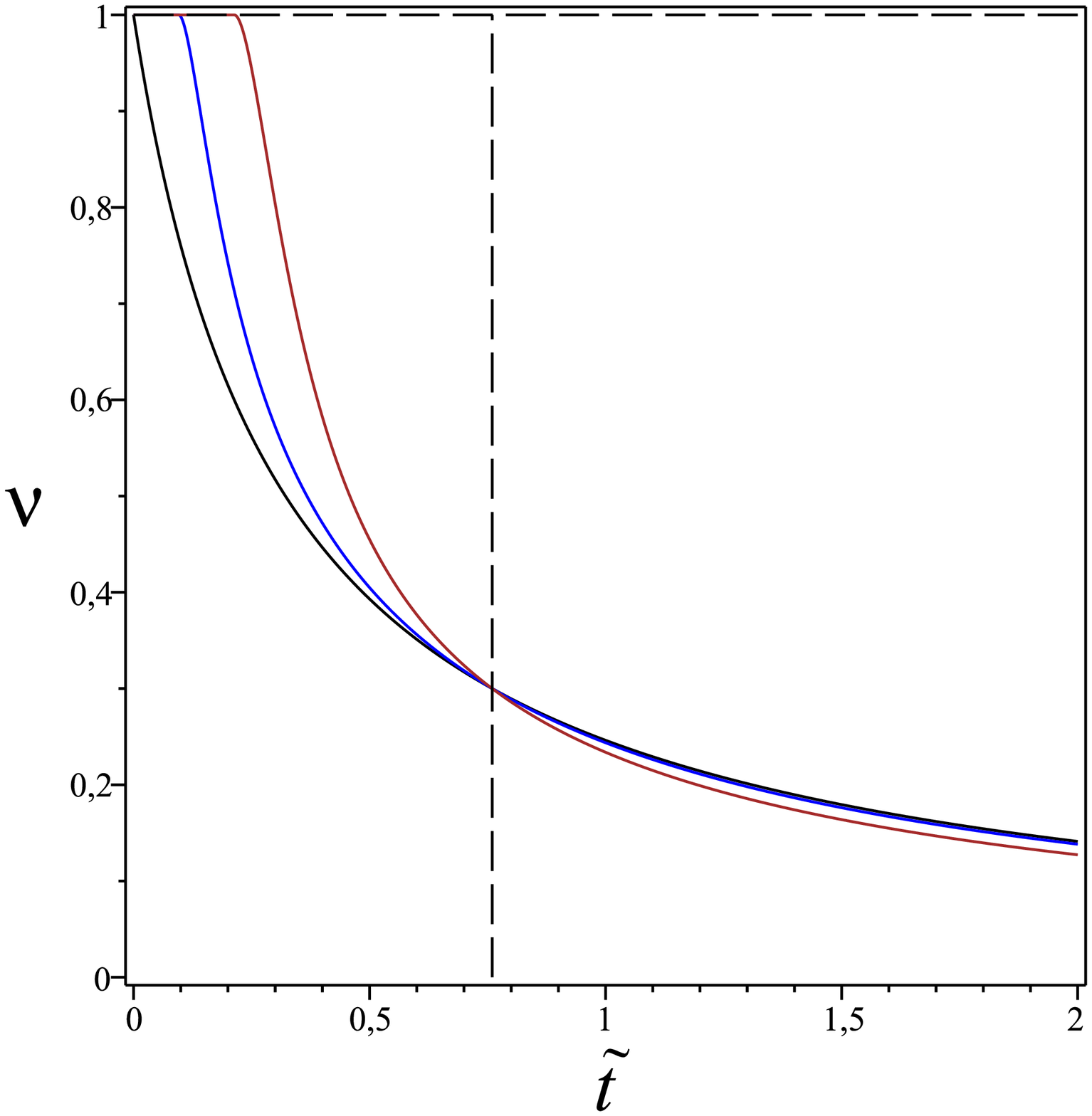}&\quad
\includegraphics[scale=0.3]{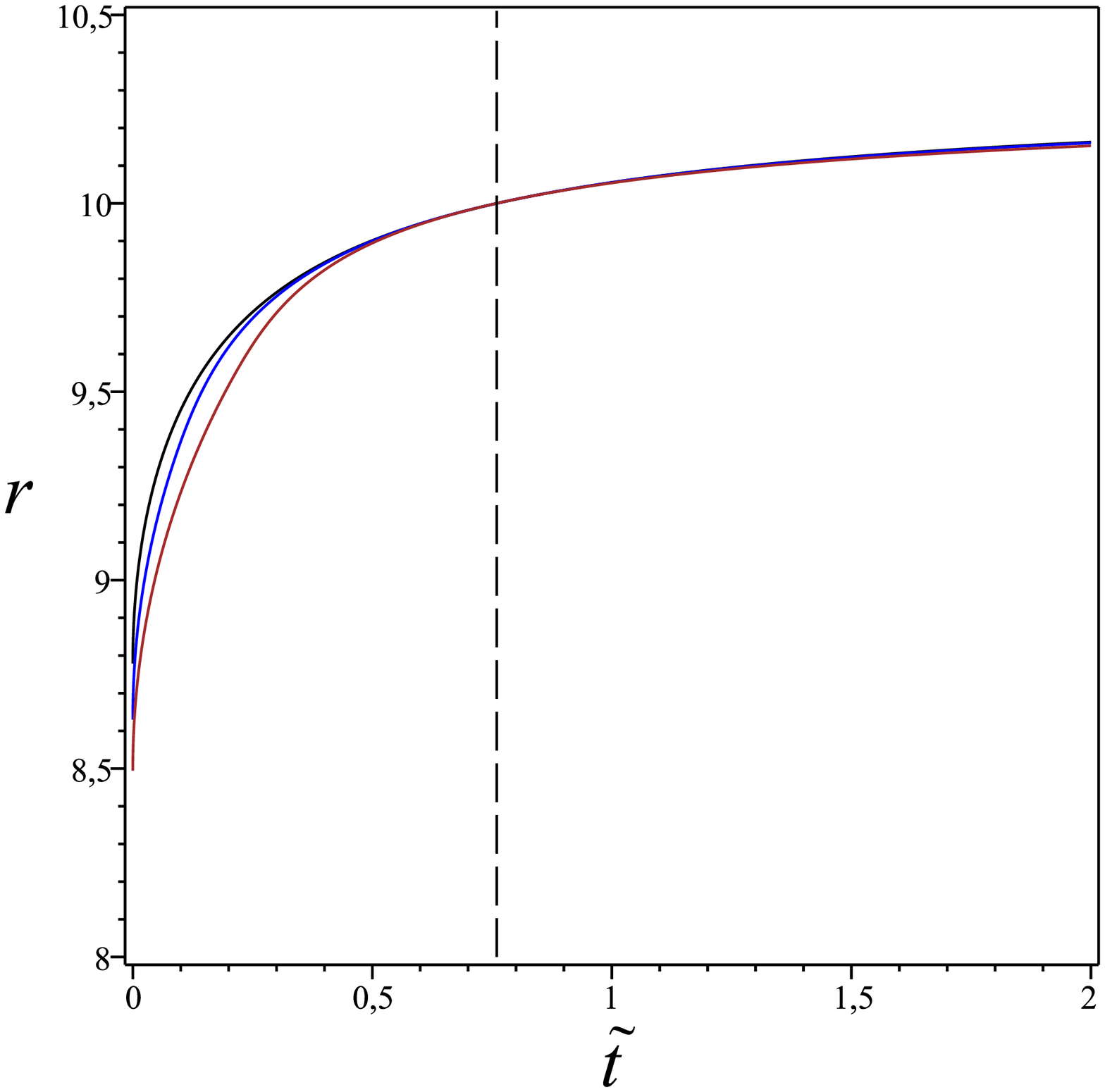}\\
\quad\mbox{(c)}\quad[w=1/3]\quad &\quad \mbox{(d)}\quad[w=1/3]\\[.4cm]
\includegraphics[scale=0.3]{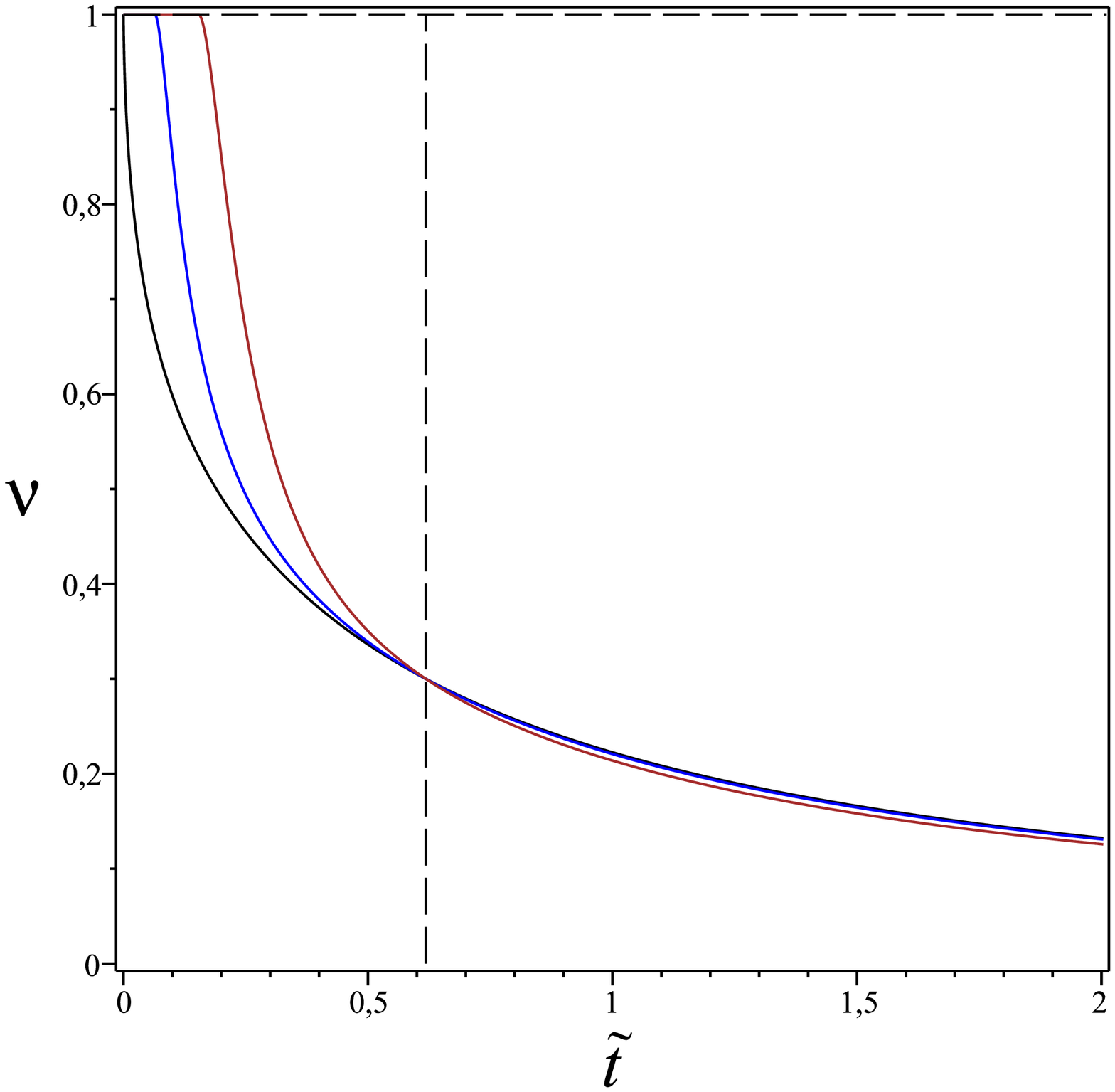}&\quad
\includegraphics[scale=0.3]{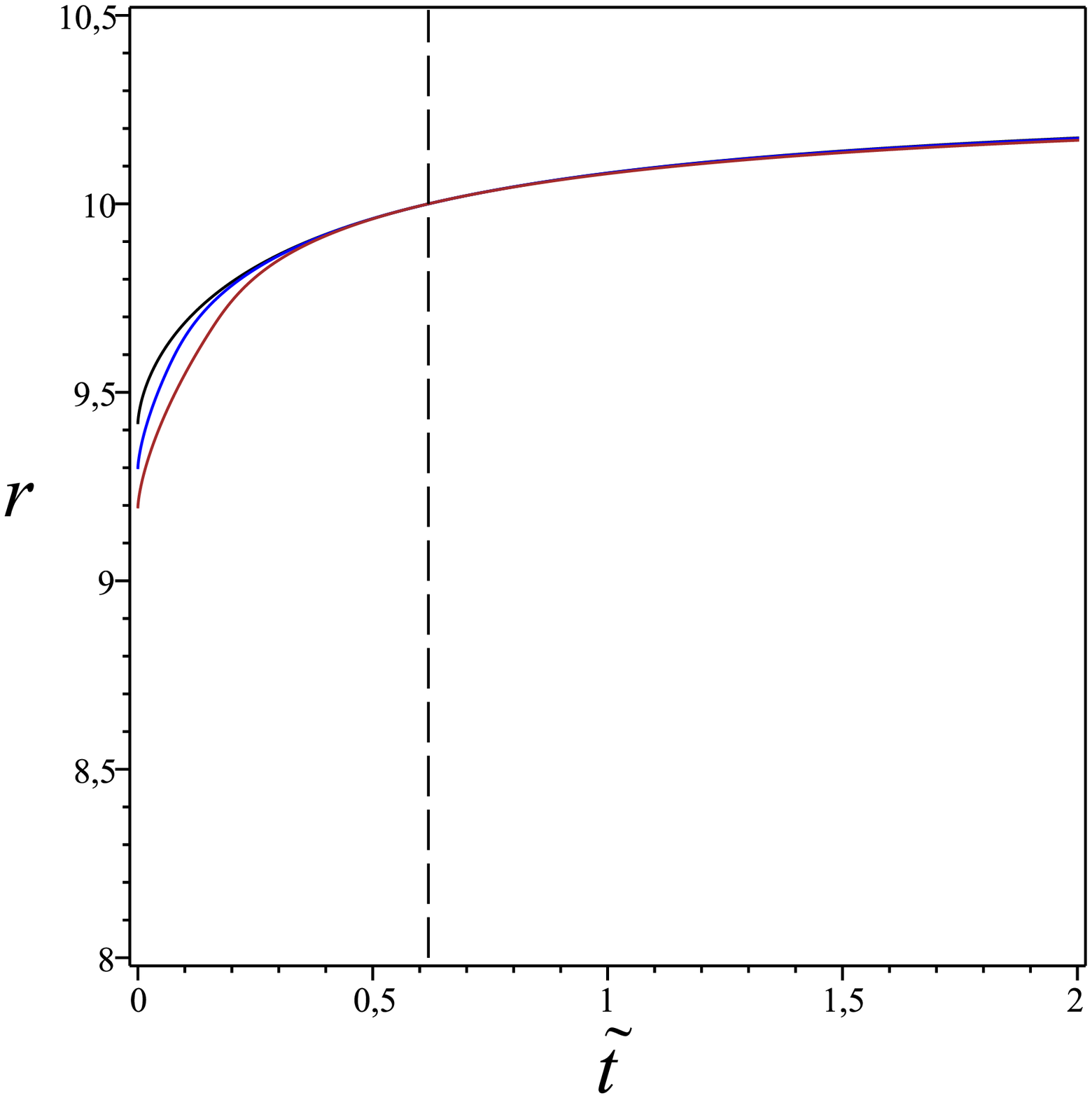}\\
\quad\mbox{(e)}\quad[w=1]\quad &\quad \mbox{(f)}\quad[w=1]\\
\end{array}$
\end{center}
\caption{[Open universe]
The behaviors of the linear velocity $\nu$ and the coordinate $r$ as functions of the time parameter $\tilde t=H_0t$ are shown for an open FRW universe for selected values of $w=[0,1/3,1]$ and different values of the friction parameter
$\tilde A=[0,1,5]$.
The set of equations (\ref{openfin})--(\ref{eqPsi2}) has been numerically integrated with $q_0=0.1$ and initial conditions $\nu_0=\nu(\Psi_0)=0.3$, $r_0=r(\Psi_0)=10$ and $\tilde t_0=\tilde t(\Psi_0)$, the latter depending on the choice of $w$.
Panels (a) and (b) refer to $w=0$, so that $\Psi_0\approx2.887$ and $\tilde t_0\approx0.846$.
Panels (c) and (d) refer to $w=1/3$, so that $\Psi_0\approx3.637$ and $\tilde t_0\approx0.760$.
Panels (e) and (f) refer to $w=1$, so that $\Psi_0\approx4.357$ and $\tilde t_0\approx0.618$.
Different values of $\tilde A$ label the various curves of panels (c)--(f) as in panels (a) and (b).
}
\label{fig:2}
\end{figure*}

The behaviors of $\nu$ and $r$ as functions of $\tilde t$ are shown in Fig. \ref{fig:2} for selected values of $w=[0,1/3,1]$ and different values of the friction parameter.
The friction force dominates at early times only, where the behavior of peculiar velocities is similar for both geodesic and accelerated cases.
Therefore, the friction parameter seems to play a minor role in this case. 
The numerical integration for $w=1/3$ and $w=1$ does not show any significant difference with respect to the case $w=0$.


\subsection{{\it Flat universe}}

The analysis of this case closely resembles that of the open universe, so that we will omit further details.

\section{Concluding Remarks}

We have investigated the dynamics of massive test particles undergoing 
friction effects in FRW spacetimes, with special emphasis 
on deviation of the particle trajectory from geodesic motion. 
The inclusion of a (weak) friction force term in the particle equations of motion is motivated by the possibility that non-gravitational interactions
 (e.g., collisions with the background fluid component) may play a role in the particle dynamics.
The background fluid is assumed to be described by a general equation of state, 
specifying its nature, e.g., as baryonic matter, radiation, dark energy.
For illustrative purposes, we have analyzed in detail the cases of pressureless matter, stiff matter and radiation, but 
our analysis is completely general and allows to investigate all of the aforementioned
fluid sources as well as mixtures thereof.
Numerical integration of the equations of motion shows major deviations from geodesic 
motion at a late stage of the evolution of a closed universe, where 
particles undergoing friction force tend to a rest state, in stark contrast to the geodesic
case, characterized by ultrarelativistic values of the particle speed.  
Backward time integration shows instead a common behavior for all 
cosmological models, i.e., both geodesic and accelerated particles approach the ultrarelativistic regime. 
Finally, it is interesting to note that the increased accuracy in recent measurements of large-scale peculiar velocities of galaxy clusters \cite{kashlinsky,dai,ma} may allow 
experimental estimates of the friction parameter.
The present work opens up several directions for future research, such as the study of
inhomogeneous cosmological models, cosmological fluids with non-ideal equations of state
$p(\rho) = w(\rho) \rho$, as well as extended theories of gravity, including extra-scalar fields.

\begin{acknowledgements}
DG is supported by the Erasmus Mundus Joint Doctorate Program by Grant Number 2011-1640 from 
the EACEA of the European Commission.
\end{acknowledgements}

\end{document}